\documentclass{article}
\usepackage{geometry}
\usepackage{soul}
\usepackage{amssymb}
\usepackage{graphicx}
\usepackage{authblk}
\usepackage[utf8]{inputenc}
\usepackage{mathptmx}
\usepackage{amsmath}
\usepackage{epsfig,amsopn}
\usepackage{graphicx}
\usepackage{braket}
\usepackage{float}
\usepackage{enumerate}
\usepackage[normalem]{ulem}
\usepackage{todonotes}

\begin{document}

\title{Entropy of Sharp Restart}

\author{Iddo Eliazar\thanks{%
E-mail: \emph{eliazar@tauex.tau.ac.il}} \and Shlomi Reuveni\thanks{%
School of Chemistry, Tel-Aviv University, 6997801, Tel-Aviv, Israel.}\thanks{%
Center for the Physics and Chemistry of Living Systems. Tel Aviv University,
6997801, Tel Aviv, Israel.}\thanks{%
The Sackler Center for Computational Molecular and Materials Science, Tel
Aviv University, 6997801, Tel Aviv, Israel.}}
\maketitle

\begin{abstract}
\ 

Restart has the potential of expediting or impeding the completion times of general random processes. Consequently, the issue of mean-performance takes center stage: quantifying how the application of restart on a process of interest impacts its completion-time’s mean. Going beyond the mean, little is known on how restart affects stochasticity measures of the completion time. This paper is the first in a duo of studies that address this knowledge gap via: a comprehensive analysis that quantifies how sharp restart -- a keystone restart protocol -- impacts the completion-time's Boltzmann-Gibbs-Shannon entropy. The analysis establishes closed-form results for sharp restart with general timers, with fast timers (high-frequency resetting), and with slow timers (low-frequency resetting). These results share a common structure: comparing the completion-time’s hazard rate to a flat benchmark -- the constant hazard rate of an exponential distribution whose entropy is equal to the completion-time’s entropy. In addition, using an information-geometric approach based on Kullback-Leibler distances, the analysis establishes results that determine the very existence of timers with which the application of sharp restart decreases or increases the completion-time’s entropy. Our work sheds first light on the intricate interplay between restart and randomness -- as gauged by the Boltzmann-Gibbs-Shannon entropy.

\bigskip\ 


\bigskip\ 

\end{abstract}

\newpage

\section{\label{1}Introduction}

\bigskip

Pioneered by Boltzmann \cite{Bol}, Gibbs \cite{Gib}, and Shannon \cite{Sha}, entropy is an elemental measure of randomness that is foundational in statistical physics and in information theory \cite{Cov}-\cite{Set}. Since its inception, entropy amassed numerous uses in a host of fields, e.g.: astrophysics \cite{Bek}, ecology \cite{Har}, mechanical engineering \cite{DY}, evolution \cite{BW}, and neural networks \cite{GML}. Yet, despite its prevalence in science and engineering, entropy is relatively under explored -- to date -- in the context of first-passage.

First-passage times (FPTs) occur naturally e.g., when considering: the first time a tracer particle reaches a certain target zone; the first time a foraging animal finds food; the first time a stock hits a certain price level; and the first time a chemical reaction occurs between two molecules. The research literature on FPTs is vast, e.g. \cite{Che2003}-\cite{Zun2022}, and we refer the reader to excellent reviews \cite{FPR1}-\cite{FPR2} and books \cite{FPB1}-\cite{FPB3} on this topic. 

Generally, one can envisage a FPT of interest as the time it takes an underlying stochastic process to complete a preset task. The FPT's entropy is intimately related to the FPT's mean -- a widely applied measure that quantifies how long, on average, it takes the process to complete its task \cite{FPB1}-\cite{God2016}. Indeed, a principal entropy-maximization result asserts that \cite{Cov}: among all non-negative random variables with a given positive mean $\mu$, the one that attains maximal entropy is Exponentially distributed. Consequently, calculating the entropy of the Exponential distribution yields the following universal entropy bound for the mean:
\begin{equation}
\mu \geq  \frac{1}{e} \exp{(\eta)} \text{ ,}  \label{101}
\end{equation}
where $\eta$ is the entropy. Hence, high entropy always implies a large mean, and a small mean always implies low entropy. In particular, the entropy bound of Eq. (\ref{101}) holds for any FPT with a positive mean.

The inherent randomness of a FPT of interest can be measured via entropy, as well as by the standard deviation. While these are markedly different measures of randomness -- the former information-based, and the latter geometry-based -- they are intimately related. Indeed, yet another principal entropy-maximization result asserts that \cite{Cov}: among all random variables with a given positive standard deviation $\sigma$, the one that attains maximal entropy is Normally distributed \cite{Cov}. Consequently, calculating the entropy of the Normal distribution yields the following universal entropy bound for the standard deviation:
\begin{equation}
\sigma \geq \frac{1}{\sqrt{2 \pi e}} \exp{(\eta)} \text{ ,} \label{102}
\end{equation}
where $\eta$ is the entropy. Hence, high entropy always implies a large standard deviation, and a small standard deviation always implies low entropy. In particular, the entropy bound of Eq. (\ref{102}) holds for any FPT with a positive standard deviation.

Recent advances in stochastic thermodynamics revealed that entropy production and fluxes can also be used to bound the means and the standard deviations of FPTs \cite{TUR1}-\cite{TUR5}. However, these thermodynamic bounds should not be confused with the entropy bounds of Eqs. (\ref{101}) and (\ref{102}) -- which manifest universal relations between the following statistics of a general random variable: its mean $\mu$ and its standard deviation $\sigma$ on the one hand, and its entropy $\eta$ on the other hand. In particular, these universal relations apply to any FPT of interest.

Wide and substantial research on FPTs took place in recent years. In particular, significant scientific work was done on the topic of \emph{first-passage under restart} \cite{Eli2007}-\cite{Yin2022}. Restart is a scheme which has the potential of expediting/impeding FPTs -- and, more broadly, task-completion durations -- of general random processes \cite{Pal2017}-\cite{Eva2020}. Indeed, the `task' of a process of interest can be a certain target zone that the process is expected to reach, and then the task-completion duration is the FPT to the target. When a restart protocol is applied to a given random process, it acts as follows: as long as the process does not accomplish its task, the protocol occasionally resets the process; the resetting is done repeatedly, till the task is accomplished. The application of restart protocols has a dramatic effect on task-completion durations: restart affects their statistical distributions, and consequently it alters their means, their variances, and their entropies.

Given the statistical distribution of interest, one often strives to `separate the wheat from the chaff' by focusing on fundamental measures that capture the distribution's key features. This goal is attained, by and large, by analyzing three principal aspects of the distribution under consideration. The first aspect is \emph{mean behavior}: the distribution's average. The second aspect is \emph{stochasticity}: the distribution's statistical heterogeneity -- its degree of inherent randomness. The third aspect is \emph{tail behavior}: the distribution's likelihood of exhibiting `rare events' -- exceptionally small and exceptionally large outcomes \cite{Asm}-\cite{Tal}.

In restart research, many scientific works addressed the mean-behavior aspect \cite{Kus2014}-\cite{Bon2021}: understanding how restart affects, on average, task-completion durations. The mean-behavior investigations established the central role of \emph{sharp restart} -- restart protocols that reset periodically, i.e. with a fixed time-lapse between consecutive resets \cite{Pal2016}-\cite{MP2SR}. In general, restart protocols can use any positive-valued random variable as their generic time-lapse between consecutive resets. The centrality of sharp restart is due to the following key fact \cite{Pal2017}-\cite{Che2018}: with regard to mean-behavior, sharp restart can out-perform any other restart protocol. Namely, if a given restart protocol attains a certain reduction/increase of the mean completion time, then: there exists a sharp restart protocol that can, at least, match this reduction/increase.
 
Following the discovery of the centrality of sharp restart, a comprehensive mean-behavior exploration of sharp restart was carried out in the duo \cite{MP1SR}-\cite{MP2SR}. Also, a comprehensive tail-behavior exploration of sharp restart was carried out in \cite{TBSR}. The studies \cite{MP1SR}-\cite{TBSR} established sets of universal results that determine and quantify the effect of sharp restart on the means and on the tails of completion-time distributions. It should be stressed that these universal results are well applicable even when information is limited, i.e.: when only partial information is provided regarding the statistical distribution (without resetting) of a completion-time under consideration.

With the mean and tail behaviours of sharp restart thoroughly investigated in \cite{MP1SR}-\cite{TBSR}, we now address the stochasticity aspect: understanding how sharp restart effects the randomness of task-completion durations. We do so in a duo of research papers. This paper, the duo's first part, gauges stochasticity via the \emph{Boltzmann-Gibbs-Shannon entropy} \cite{Bol}-\cite{Sha} -- henceforth termed, in short, \emph{entropy}. This paper presents a comprehensive, entropy-based, stochasticity analysis of sharp restart. In turn, the analysis establishes a set of universal results that determine and quantify the effect of sharp restart on the entropies of completion-time distributions. The duo's second part shall gauge stochasticity via \emph{diversity} -- a profound measure of randomness that is widely applied in ecology \cite{Hil}-\cite{LL}.

The remainder of this paper is organized as follows. Section \ref{1} reviews sharp restart from an algorithmic perspective: the algorithm's input is the task-completion duration of a random process of interest; and the algorithm's output is the task-completion duration that is attained by applying sharp-restart to the underlying process. Section \ref{8} establishes, in terms of the input's statistics, a general closed-form formula for the output's entropy; based on the general formula, this section presents a preliminary analysis of the effect sharp-restart has on entropy. Section \ref{3} re-formulates the results of the previous section in terms of the input's hazard rate, providing a neat representation of the output's entropy; based on the hazard-rate reformulations, this section presents general criteria that determine -- for any given timer -- the effect sharp-restart has on entropy. Section \ref{4} addresses the special cases of high-frequency and low-frequency resetting, and shows that: for these cases, the effect of sharp-restart on entropy is determined by the limit-values of the input's hazard rate at zero and at infinity. Section \ref{5} analyzes sharp restart from yet another perspective: the very existence of timers with which sharp restart decreases/increases entropy. To that end section \ref{5} employs an information-geometry approach -- measuring Kullback-Leibler distances between density functions that are induced by the input's statistics. Section \ref{6} concludes with a summary of the key results that were established in this paper, and with an outlook towards the second part of this duo: the diversity of sharp restart. To ease the follow of reading, the derivations of key results are detailed in the Methods.

\section{\label{1}Sharp restart}

\emph{Sharp restart} is an \emph{algorithm} that is described as follows \cite{MP1SR}-\cite{TBSR}. There is a general task with completion time $T$, a positive-valued random variable. To this task a three-steps algorithm, with a positive deterministic timer $\tau $, is applied. Step I: initiate simultaneously the task and the timer. Step II: if the task is accomplished up to the timer's expiration -- i.e. if $T\leq \tau $ -- then stop upon completion. Step III: if the task is not accomplished up to the timer's expiration -- i.e. if $T>\tau $ -- then, as the timer expires, go back to
Step I.

The sharp-restart algorithm generates an iterative process of independent and statistically identical task-completion trials. This process halts during its first successful trial, and we denote by $T_{R}$ its halting time. Namely, $T_{R}$ is the overall time it takes -- when the sharp-restart algorithm is applied -- to complete the task. The sharp-restart algorithm is a \emph{non-linear mapping} whose input is the random variable $T$, whose output is the random variable $T_{R}$, and whose parameter is the deterministic timer $\tau $.

Henceforth, we set the task-completion process to start at time $t=0$; thus, the process takes place over the non-negative time axis $t\geq 0$. Along this paper we use the following notation regarding the input's statistics: $F\left( t\right) =\Pr \left( T\leq t\right) $ ($t\geq 0$) denotes the distribution function; $\bar{F}\left( t\right) =\Pr \left( T>t\right) $ ($t\geq 0$) denotes the survival function; and $f\left( t\right) =F^{\prime }\left(t\right) =-\bar{F}^{\prime }\left( t\right) $ ($t>0$) denotes the density function. The input's density function is considered to be positive-valued over the positive half-line: $f\left( t\right) >0$ for all $t>0$.

As established in \cite{TBSR}: in terms of the input's survival and density functions, the output's density function $f_{R}\left( t\right) $ admits the following representation \cite{TBSR}:
\begin{equation}
f_{R}\left( \tau n+u\right) =\bar{F}\left( \tau \right) ^{n}f\left( u\right) 
\text{ ,}  \label{201}
\end{equation}
where $n=0,1,2,\cdots $, and where $0\leq u<\tau $. Indeed, in order that the output $T_{R}$ be realized at time $t=\tau n+u$ we need that: \textbf{A}) the first $n$ task-completion trials be unsuccessful; and \textbf{B}) the task be accomplished right at the time epoch $u$ of the $(n+1)^{\text{th}}$ task-completion trial. Event \textbf{A} occurs with probability $\bar{F}\left( \tau \right) ^{n}$, event \textbf{B} occurs with likelihood $f\left( u\right) $, and hence: the likelihood that the output be realized right at time $t=\tau n+u$ is given by the right-hand side of Eq. (\ref{201}).

\section{\label{8} The entropy of sharp restart}

Following Boltzmann, Gibbs and Shannon, the entropy of a general real-valued random variable $X$ is:
\begin{equation}
\mathcal{E}\left[ X\right] =-\int_{-\infty }^{\infty } \ln \left[ \varphi \left( x\right) \right] \varphi \left(x\right) dx\text{ ,}
\end{equation}
where $\varphi \left(x\right) $ ($-\infty <x<\infty $) is the probability density function of the random variable $X$. The entropy $\mathcal{E}[X]$ can be interpreted as the weighted average of the function $\ln{[1 / \varphi(x)]}$, where the weights are given by the density function $\varphi (x)$. Alternatively, the entropy $\mathcal{E}[X]$ can be interpreted as the mean of the random variable $\ln{[1 / \varphi(X)]}$.

The goal of this paper is to explore the effect of the sharp-restart algorithm on entropy. To that end we denote by $\eta =\mathcal{E}\left[ T\right] $ the input's entropy, and consider it to be finite $-\infty <\eta <\infty $. Also, we denote by $E\left( \tau \right) =\mathcal{E}\left[ T_{R}\right] $ the output's entropy; this notation underscores the fact that the output's entropy is a function of the timer $\tau $, the algorithm's parameter. And, we use the following terminology:
\begin{enumerate}
\item[$\bullet $] Sharp restart with timer $\tau $ \emph{decreases entropy} if the output's entropy is smaller than the input's entropy, $E\left( \tau \right) <\eta $.

\item[$\bullet $] Sharp restart with timer $\tau $ \emph{increases entropy} if the output's entropy is larger than the input's entropy, $E\left( \tau \right) >\eta $.
\end{enumerate}

\begin{figure}[t!]
\centering
\includegraphics[width=13.25cm]{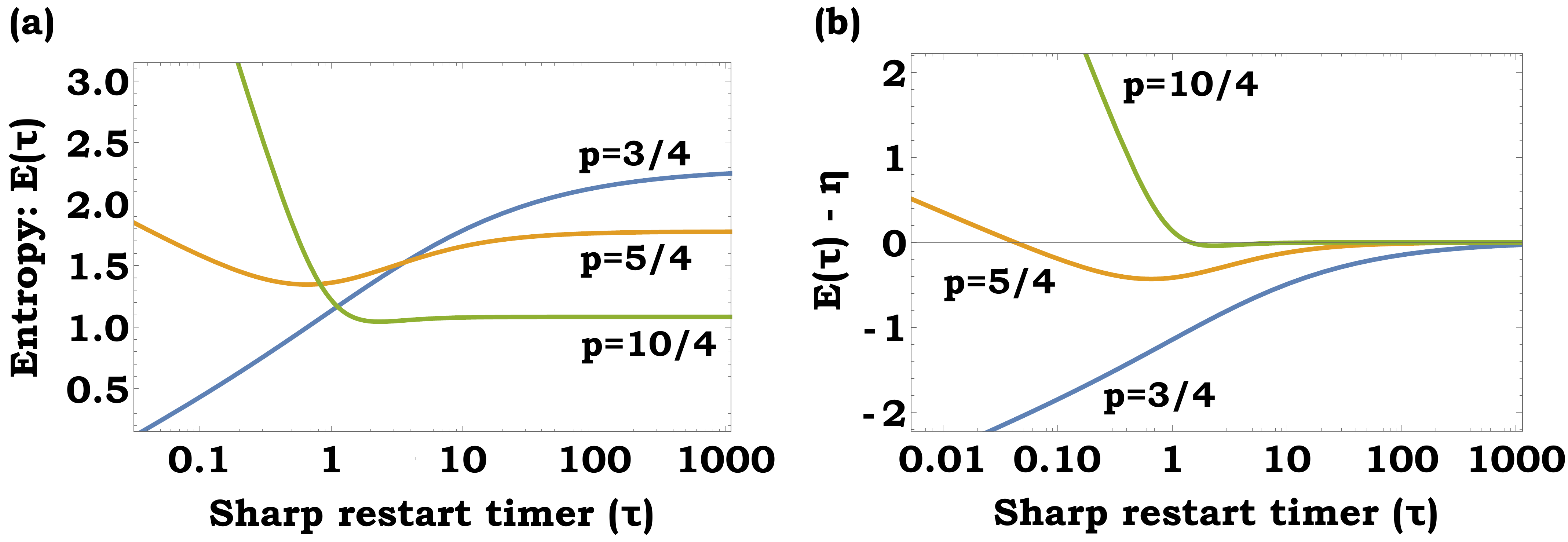}
\caption{Entropy of sharp restart -- Pareto example. Consider the input statistics to be Pareto type III. Namely, the input's survival function is $\bar{F}(t)=1/(1+t^p)$, where $p$ is a positive power. Using Eq. (\ref{202}), the output’s entropy $E\left( \tau \right)$ is plotted vs. its variable $\tau$, the sharp-restart timer (panel a). Similarly, the output-input entropy difference $E\left( \tau \right) - \eta$ is plotted vs. the variable $\tau$ (panel b). Evidently, the behaviors of the output's entropy and of the entropy difference are highly sensitive to the Pareto power $p$, as well as to the sharp-restart timer $\tau$. Note that there are values of the Pareto power for which: sharp restart with some timers decreases entropy, whereas sharp restart with other timers increases entropy. Thus, even from this simple example -- which has only a single parameter -- we learn that the effect of sharp restart on entropy can be intricate (and highly non-trivial)}.
\label{KLFig}
\end{figure}

Computing the output's entropy from the output's density function by use of Eq. (\ref{201}), we establish that
\begin{equation}
E\left( \tau \right) =-\frac{1}{F\left( \tau \right) }\left\{ \bar{F}\left(\tau \right) \ln \left[ \bar{F}\left( \tau \right) \right] +\int_{0}^{\tau}f\left( t\right) \ln \left[ f\left( t\right) \right] dt\right\} \text{ .}
\label{202}
\end{equation}
With Eq. (\ref{202}) at hand, we further establish that the difference between the output's and the input's entropies is
\begin{equation}
E\left( \tau \right) -\eta =\frac{\bar{F}\left( \tau \right) }{F\left( \tau \right) }\left\{ \eta -\ln \left[ \bar{F}\left( \tau \right) \right] +\frac{1%
}{\bar{F}\left( \tau \right) }\int_{\tau }^{\infty }f\left( t\right) \ln \left[ f\left( t\right) \right] dt\right\} \text{ .}  \label{203}
\end{equation}
The derivations of Eqs. (\ref{202}) and (\ref{203}) are detailed in the Methods.

Equipped with Eq. (\ref{202}), and given the statistical distribution of an input of interest, one can explore the effect of sharp restart on entropy. To demonstrate the potential wealth of scenarios, consider the following example: an input whose statistics are Pareto type III; these statistics are also known as log-logistic. The input's survival function is then $\bar{F}(t)=1/(1+t^p)$, where $p$ is a positive power. Using Eq. (\ref{202}), the output's entropy $E\left( \tau \right)$ and the output-input entropy difference $E\left( \tau \right) - \eta$ are plotted against the sharp-restart timer $\tau$ (Fig. 1). Even for this simple example -- which has only a single parameter -- it is evident that: the output's entropy and the entropy difference can display highly non-trivial behaviors with respect to the sharp-restart timer $\tau$, and with respect to the parameters of the input's statistics (here the Pareto power $p$).

Eq. (\ref{203}) straightforwardly yields the following pair of criteria that determine if the application of the sharp-restart algorithm decreases or increases entropy.

\begin{enumerate}
\item[$\bullet $] If 
\begin{equation}
\eta < \ln \left[ \bar{F}\left( \tau \right) \right] - \frac{1}{\bar{F}\left(\tau \right) }\int_{\tau }^{\infty }f\left( t\right) \ln \left[ f\left(t\right) \right] dt  \label{204}
\end{equation}
then sharp restart with the timer $\tau$ decreases entropy.

\item[$\bullet $] And, if 
\begin{equation}
\eta > \ln \left[ \bar{F}\left( \tau \right) \right] - \frac{1}{\bar{F}\left(\tau \right) }\int_{\tau }^{\infty }f\left( t\right) \ln \left[ f\left(t\right) \right] dt  \label{205}
\end{equation}
then sharp restart with the timer $\tau$ increases entropy.
\end{enumerate}

The results of this section facilitate a precise quantitative analysis of sharp restart. Indeed, given an input of interest, one can plug the input's survival and density functions into the closed-form formulae of this section, and compare -- in most cases numerically -- the output's entropy to the input's entropy. This comparison was demonstrated via the Pareto type III example above (Fig. 1). However, this comparison is done on a case-by-case basis, and hence it is neither very practical nor very insightful. Moreover, this comparison is feasible when the input's statistics are known in full detail, and it is not feasible when only partial information regarding the input's statistics is available. To better understand the effect of sharp restart on entropy, we shall now continue the exploration -- doing so by using the notion of hazard rate (in sections \ref{3} and \ref{4}), and by using the notion of relative entropy (in section \ref{5}).

\section{\label{3} Hazard-rate approach to sharp-restart entropy}

The input's \emph{hazard function} plays a focal role in the mean-performance analysis \cite{MP1SR}, as well as in the tail-behavior analysis \cite{TBSR}, of the sharp-restart algorithm. This hazard function is the negative logarithmic derivative of the input's survival function:
\begin{equation}
H\left( t\right) =-\left\{ \ln \left[ \bar{F}\left( t\right) \right] \right\} ^{\prime }=\frac{f\left( t\right) }{\bar{F}\left( t\right) }\text{.}  \label{300}
\end{equation}
The hazard function has the following probabilistic meaning \cite{MP1SR},\cite{TBSR}: $H\left( t\right)$ is the likelihood that the input be realized right after time $t$, given the information that the input was not realized up to time $t$. The hazard function -- a.k.a. \textquotedblleft hazard rate\textquotedblright\ and \textquotedblleft failure rate\textquotedblright\ -- is a widely applied tool in survival analysis \cite{KP}-\cite{Col}, and in reliability engineering \cite{BP}-\cite{Dhi}.

In terms of its hazard and density functions, the input's entropy admits the representation
\begin{equation}
\eta = \int_{0}^{\infty} \ln{ \left[ \frac{e}{H(t)} \right] }f(t)dt \text{ .}
\label{3001}
\end{equation}
Namely, the input's entropy $\eta$ is the weighted average of the function $\ln{[e/H(t)]}$, where the weights are given by the input's density function $f(t)$. Described probabilistically, the entropy $\eta$ is the expectation of the random variable $\ln{ [e / H(T)]}$. The derivation of Eq. (\ref{3001}) is detailed in the Methods.

The entropy representation appearing in Eq. (\ref{3001}) is general, i.e. it applies to any positive-valued random variable with a density function. In particular, this representation applies to the output of the sharp-restart algorithm. As shown in \cite{TBSR}, it follows from Eq. (\ref{201}) that the output's hazard rate has a neat and compact form: a periodic concatenation of the input's hazard function over the temporal interval $(0,\tau)$. In turn, this periodic form of the output's hazard rate, together with the output's density function of Eq. (\ref{201}), yield a neat representation of the output's entropy -- which we shall now present.

The representation of the output's entropy involves the input's hazard function $H(t)$, as well as the following function: $f_{\tau}(t)=f(t)/F(\tau)$ ($0<t<\tau$), which is the conditional density function of the input $T$ -- given the information that the input is no-larger than the timer, $T \leq \tau$. In terms of these two functions, the representation is
\begin{equation}
E(\tau) = \int_{0}^{\tau} \ln{ \left[ \frac{e}{H(t)} \right] } f_{\tau}(t) dt \text{ .}
\label{3002}
\end{equation}
Namely, the output's entropy $E(\tau)$ is the weighted average of the function $\ln{[e/H(t)]}$, where the weights are given by the conditional density function $f_{\tau}(t)$. Described probabilistically, the entropy $E(\tau)$ is the conditional expectation -- with respect to the information $T \leq \tau$ -- of the random variable $\ln{ [e / H(T)]}$. The derivation of Eq. (\ref{3002}) is detailed in the Methods.

The entropy representations of Eq. (\ref{3001}) and Eq. (\ref{3002}) share a common pattern: a weighted average of the function $\ln{[e/H(t)]}$. The two representations differ by the weights they use: the input's density function $f(t)$ in the former, and the conditional density function $f_{\tau}(t)$ in the latter. Note that in the timer limit $\tau \rightarrow \infty$ the conditional density function $f_{\tau}(t)$ converges to the input's density function $f(t)$, and the output's entropy $E(\tau)$ converges to the input's entropy $\eta$.

Equations (\ref{202}) and (\ref{3002}) present equivalent formulations of the output's entropy. However, as noted at the end of section \ref{8}, the formulation of Eq. (\ref{202}) is not very amicable to work with, and it is not so easy to deduce insights from this formulation. On the other hand, as we shall show below and in the next section, Eq. (\ref{3002}) is a practical `working tool', and deep and useful insights will be drawn from it.

\subsection{Entropic invariance}

A key insight that emerges from Eq. (\ref{3002}) regards entropic invariance -- the scenario in which entropy is invariant with respect to the application of the sharp-restart algorithm. Namely, entropic invariance is characterized by a `flat' entropy function of the output: $E( \tau)=\eta$ for all timers $\tau$.

Substituting the flat output entropy into Eq. (\ref{3002}), and then multiplying both sides by the quantity $F(\tau)$ yields $\eta F(\tau) = \int_{0}^{\tau} \ln{ [e/H(t)] } f(t) dt$. In turn, differentiating with respect to the timer parameter $\tau$ further yields $\eta=\ln{ [e/H(t)] }$. Hence, a flat entropy function of the output, $E( \tau)=\eta$, implies a flat hazard function of the input: $H(t)=e/ \exp{(\eta)}$ for all times $t>0$. On the other hand, substituting this flat hazard function into Eq. (\ref{3002}) yields the flat output entropy from which we set off. So, we obtain that: a flat hazard function $H(t)=e/ \exp{(\eta)}$ of the input is equivalent to a flat entropy function $E( \tau)=\eta$ of the output.

With regard to positive-valued random variables, it is a well known fact that a flat hazard function characterizes the \emph{Exponential distribution}. Also, the uniform height of a flat hazard function is commonly referred to as the \emph{rate} of the corresponding Exponential distribution. Consequently, we arrive at the following conclusion: entropic invariance holds if and only if the input is Exponentially-distributed.

As noted above, the rate of an Exponentially-distributed input with entropy $\eta$ is
\begin{equation}
r_{\exp} = \frac{e}{\exp{(\eta)}}.
\label{3010}
\end{equation}
The specific rate that is distinguished in Eq. (\ref{3010}) will appear time and again in various results that we shall establish below. The denominator in the rate of Eq. (\ref{3010}) -- the exponentiation of the input's entropy -- is commonly refereed to as the input's \emph{perplexity} \cite{JMB}. 

\subsection{Hazard-rate criteria based on left-tail statistics}

Subtracting the input's entropy $\eta$ from both sides of Eq. (\ref{3002}), and using the specific rate $r_{\exp}$ of Eq. (\ref{3010}), a bit of algebra yields the following formula for the difference between the output's and input's entropies:
\begin{equation}
E\left( \tau \right) -\eta =\frac{1}{F\left( \tau \right) }\int_{0}^{\tau }\ln \left[ \frac{r_{\exp}}{H\left( t\right) }\right] f\left( t\right) dt\text{.}
\label{301}
\end{equation}
As noted above, the rate $r_{\exp}$ manifests the flat hazard function that characterizes an Exponential distribution whose entropy is the input's entropy $\eta$. So, in Eq. (\ref{301}), the rate $r_{\exp}$ serves as an `Exponential hazard-function benchmark' to which the input's hazard function $H(t)$ is compared. The derivation of Eq. (\ref{301}) is detailed in the Methods.

As the input's distribution function is positive-valued, Eq. (\ref{301}) implies the following pair of criteria that determine if the application of the sharp-restart algorithm decreases or increases entropy.

\begin{enumerate}
\item[$\bullet $] If 
\begin{equation}
\int_{0}^{\tau}\ln \left[ \frac{r_{\exp}}{H\left( t\right)}\right] f\left(t\right) dt<0  \label{305}
\end{equation}
then sharp restart with the timer $\tau $ decreases entropy. 

\item[$\bullet $] And, if  
\begin{equation}
\int_{0 }^{\tau}\ln \left[ \frac{r_{\exp}}{H\left( t\right)}\right] f\left(t\right) dt>0  \label{306}
\end{equation}
then sharp restart with the timer $\tau $ increases entropy.
\end{enumerate}
Note that in order to use the above criteria one only needs knowledge of the left-tail statistics of the input. We shall now demonstrate the criteria of Eqs. (\ref{305})-(\ref{306}) `in action'.

Consider the following Exponential-Pareto example: an input whose statistics are Exponential up to the time point $t=1$, and are Pareto afterward. Specifically, the input's density function is $f(t)=r \exp(-rt)$ over the temporal interval $0<t<1$, where $r$ is a positive rate. And, the input's density function is $f(t)= \exp(-r)pt^{-p-1}$ over the temporal ray $1<t<\infty$, where $p$ is a positive power. A calculation implies that the logarithm of the corresponding level $r_{\exp}$ is  $\ln{(r_{\exp})}=\exp(-r)[\ln{(p)}-\frac{1}{p}]+[1-\exp(-r)]\ln{(r)}$. Also, note that over the temporal interval $0<t<1$ the hazard function is flat $H(t)=r$, and hence $\ln{[r_{\exp}/H(t)]} = \exp(-r)[\ln{(p)}-\frac{1}{p}-\ln{(r)}]$. Consequently, for timers $\tau<1$, the criteria imply that: if $\ln{(r)}>\ln{(p)}-\frac{1}{p}$ then sharp restart decreases entropy; and if $\ln{(r)}<\ln{(p)}-\frac{1}{p}$ then sharp restart increases entropy.

\subsection{Hazard-rate criteria based on right-tail statistics}

Applying a bit of algebra to Eq. (\ref{301}), while using Eq. (\ref{3001}) and the specific rate $r_{\exp}$ of Eq. (\ref{3010}), yields the following formula for the difference between the output's and input's entropies:
\begin{equation}
E\left( \tau \right) -\eta =\frac{1}{F\left( \tau \right) }\int_{\tau}^{\infty }\ln \left[ \frac{H\left( t\right) }{r_{\exp}}\right] f\left( t\right) dt \text{ .}  
\label{302}
\end{equation}
As noted above, the rate $r_{\exp}$ manifests the flat hazard function that characterizes an Exponential distribution whose entropy is the input's entropy $\eta$. So, as in Eq. (\ref{301}): in Eq. (\ref{302}) the rate $r_{\exp}$ serves as an `Exponential hazard-function benchmark' to which the input's hazard function $H(t)$ is compared. The derivation of Eq. (\ref{302}) is detailed in the Methods.

As the input's distribution function is positive-valued, Eq. (\ref{302}) implies the following pair of criteria that determine if the application of the sharp-restart algorithm decreases or increases entropy.

\begin{enumerate}
\item[$\bullet $] If 
\begin{equation}
\int_{\tau }^{\infty }\ln \left[ \frac{H\left( t\right) }{r_{\exp}}\right] f\left(
t\right) dt<0  \label{303}
\end{equation}
then sharp restart with the timer $\tau $ decreases entropy.

\item[$\bullet $] And, if
\begin{equation}
\int_{\tau }^{\infty }\ln \left[ \frac{H\left( t\right) }{r_{\exp}}\right] f\left(
t\right) dt>0  \label{304}
\end{equation}
then sharp restart with the timer $\tau $ increases entropy.
\end{enumerate}
Note that in order to use the above criteria one only needs knowledge of the right-tail statistics of the input. We shall now demonstrate the criteria of Eqs. (\ref{303})-(\ref{304}) `in action'.

Consider the following Uniform-Exponential example: an input whose statistics are Uniform up to the time point $t=u$, and are Exponential afterward. Specifically, the input's density function is $f(t)=1$ over the temporal interval $0<t<u$, where $u$ is a positive number that is smaller than $1$. And, the input's density function is $f(t)=(1-u)r \exp{[-r(t-u)}]$ over the temporal ray $u<t<\infty$, where $r$ is a positive rate. A calculation implies that the logarithm of the corresponding level $r_{\exp}$ is $\ln{(r_{\exp})}=u+(1-u) \ln{[(1-u)r]}$. Also, note that over the temporal interval ray $u<t<\infty$ the hazard function is flat $H(t)=r$, and hence $\ln{[H(t)/r_{\exp}]} = u \ln{(r)} - u - (1-u) \ln{(1-u)}$. Consequently, for timers $\tau>u$, the criteria imply that: if $\ln{(r)}<1+\frac{1}{u}(1-u) \ln{(1-u)}$ then sharp restart decreases entropy; and if $\ln{(r)}>1+\frac{1}{u}(1-u) \ln{(1-u)}$ then sharp restart increases entropy.

\subsection{Conclusion}

Equation (\ref{301}) and Eq. (\ref{302}) are equivalent re-formulations of Eq. (\ref{203}). The formulae appearing in Eqs. (\ref{301}) and (\ref{302}) provide two perspectives of the difference between the output's and the input's entropies: one via the input's statistics over the temporal interval $(0,\tau)$; and one via the input's statistics over the temporal ray $(\tau,\infty)$. The former perspective yielded the criteria of Eqs. (\ref{305})-(\ref{306}), whose application was demonstrated by the Exponential-Pareto example. The latter perspective yielded the criteria of Eqs. (\ref{303})-(\ref{304}), whose application was demonstrated by the Uniform-Exponential example. We emphasize that the two pairs of criteria are equivalent. Given an input of interest, one can choose which pair of criteria is more convenient to apply (as we did in the above examples). The two perspectives that were set and employed in this section will be further employed in the next section.

\section{\label{4}Sharp restart with high- and low-frequency resetting}

In sections \ref{8} and \ref{3} we established results regarding the effect of the sharp-restart algorithm -- with a general positive timer $\tau$ -- on entropy. In this section we address the asymptotic timer limits $\tau \rightarrow 0$ and $\tau \rightarrow \infty $. These limits correspond, respectively, to the following extreme cases. \textbf{I}) The case of \emph{fast timers}: very small timers, $\tau\ll1$, which manifest high-frequency resetting. \textbf{II}) The case of \emph{slow timers}: very large timers, $\tau\gg1$, which manifest low-frequency resetting. 

The asymptotic analysis of this section will use the following shorthand notation: $\varphi \left( 0\right) =\lim_{t\rightarrow 0}\varphi \left( t\right) $ and $\varphi \left( \infty \right) =\lim_{t\rightarrow \infty }\varphi \left( t\right) $ denote the limit values of a general real-valued function $\varphi \left( t\right) $ that is defined over the positive half-line ($t>0$); these limit values are assumed to exist in the wide sense, $0\leq \varphi \left( 0\right) ,\varphi \left( \infty \right) \leq \infty $. Also, as in the previous section, in this section we will use the the specific rate $r_{\exp}$ of Eq. (\ref{3010}).

\subsection{\label{41}Fast timers}

Considering the fast-timers limit $\tau \rightarrow 0$, a straightforward calculation that applies L'Hospital's rule to Eq. (\ref{301}) yields 
\begin{equation}
E\left( 0\right) -\eta =\ln \left[ \frac{r_{\exp}}{H\left( 0\right) }\right] \text{.}  \label{412}
\end{equation}
In turn, Eq. (\ref{412}) implies the following pair of criteria that determine if the application of the sharp-restart algorithm -- for sufficiently small timers -- decreases or increases entropy.

\begin{enumerate}
\item[$\bullet $] If $H\left( 0\right) >r_{\exp}$ then sharp restart with sufficiently small timers $\tau $ decreases entropy; in particular, this criterion holds automatically when $H\left( 0\right) =\infty $.

\item[$\bullet $] And, if   $H\left( 0\right) <r_{\exp}$ then sharp restart with sufficiently small timers $\tau $ increases entropy; in particular, this criterion holds automatically when $H\left( 0\right) =0$.
\end{enumerate}

How small should the timer $\tau $ be in order to qualify as \textquotedblleft sufficiently small\textquotedblright ? To answer this question assume that the input's hazard function $H\left( t\right) $ is continuous, and consider the minimal point at which this function intersects the positive level $r_{\exp}$, i.e.: 
\begin{equation}
\tau _{\ast }=\inf \left\{ t\geq 0\text{ }|\text{ }H\left( t\right) =r_{\exp}\right\} \text{ ;}  \label{413}
\end{equation}
if there is no such intersection point then $\tau _{\ast }=\infty $. With the minimal point $\tau _{\ast }$ defined, Eq. (\ref{301}) implies that: the range of \textquotedblleft sufficiently small\textquotedblright\ timers is $0<\tau <\tau _{\ast }$.

\subsection{\label{42}Slow timers}

Considering the slow-timers limit $\tau
\rightarrow \infty $, Eq. (\ref{302}) implies that $E\left( \infty \right)=\eta $. Moreover, a straightforward calculation that applies L'Hospital's rule to Eq. (\ref{302}) yields 
\begin{equation}
\lim_{\tau \rightarrow \infty }\frac{E\left( \tau \right) -\eta }{\bar{F}\left( \tau \right) }=\ln \left[ \frac{H\left( \infty \right) }{r_{\exp}}\right] 
\text{ .}  \label{422}
\end{equation}
In turn, Eq. (\ref{422}) implies the following pair of criteria that determine if the application of the sharp-restart algorithm -- for sufficiently large timers -- decreases or increases entropy.

\begin{enumerate}
\item[$\bullet $] If $H\left( \infty \right) <r_{\exp}$ then sharp restart with sufficiently large timers $\tau $ decreases entropy; in particular, this criterion holds automatically when $H\left( \infty \right) =0$.

\item[$\bullet $] And, if   $H\left( \infty \right) >r_{\exp}$ then sharp restart with sufficiently large timers $\tau $ increases entropy; in particular, this criterion holds automatically when $H\left( \infty \right) =\infty$.
\end{enumerate}

How large should the timer $\tau $ be in order to qualify as \textquotedblleft sufficiently large\textquotedblright ? To answer this question assume that the input's hazard function $H\left( t\right) $ is continuous, and consider the maximal point at which this function intersects the positive level $r_{\exp}$, i.e.:  
\begin{equation}
\tau ^{\ast }=\sup \left\{ t\geq 0\text{ }|\text{ }H\left( t\right)=r_{\exp}\right\} \text{ ;}  \label{423}
\end{equation}
if there is no such intersection point then $\tau ^{\ast }=0$. With the maximal point $\tau ^{\ast }$ defined, Eq. (\ref{302}) implies that: the range of \textquotedblleft sufficiently large\textquotedblright\ timers is $\tau ^{\ast }<\tau <\infty $.

\bigskip

\section{\label{5}Kullback-Leibler analysis of sharp-restart entropy}

In the previous sections we explored sharp-restart with specific timers: general positive timers (sections \ref{8} and \ref{3}), and fast and slow timers (section \ref{4}). In this section we present a different approach and establish \emph{existence results}: criteria that determine the very existence of timers with which the application of the sharp-restart algorithm decreases or increases entropy. 

The existence results use four different density functions that are `induced' by the input's density function. To these four density functions (as well as to the input's density function), the existence results apply relative entropy -- a.k.a. the \emph{Kullback-Leibler divergence} \cite{KL}-\cite{Kul} -- which is of key importance in information theory \cite{Cov}. This section is organized as follows: we introduce the four density functions; then we briefly review the notion of relative entropy; and thereafter we present the existence results.

\subsection{\label{50} Four densities}

As noted above, the existence results involve four density functions that are induced by the input's density function $f(t)$ ($t>0$). As the input's density function, all four density functions below are also defined over the positive half-line ($t>0$). Along this section we consider the input's mean -- henceforth denoted $\mu $ -- to be positive (i.e. it is neither zero, nor is it infinite).

Two density functions emanate from the renewal process that is generated by the input \cite{Cox}-\cite{Ros}. Specifically, the renewal process is a sequence of temporal \textquotedblleft renewal epochs\textquotedblright\ $T_{1},T_{1}+T_{2},T_{1}+T_{2}+T_{3},\cdots $, whose \textquotedblleft inter-renewal periods\textquotedblright\ $\left\{ T_{1},T_{2},T_{3},\cdots \right\} $ are independent and identically distributed copies of the input $T$.

Standing at the positive time point $t_{0}$, and looking forward in time, consider the waiting duration till the first renewal epoch after the time point $t_{0}$. This waiting duration converges in law, as $t_{0}\rightarrow \infty $, to a limiting random variable that is termed the \emph{residual lifetime} of the input $T$. The density function of the residual lifetime is \cite{Cox}-\cite{Ros}:
\begin{equation}
f_{res}\left( t\right) =\frac{1}{\mu }\bar{F}\left( t\right) \text{ ,}
\label{501}
\end{equation}
($t>0$). The residual lifetime played a principal role in the mean-performance analysis of sharp restart \cite{MP1SR}, as well as in the mean-performance analysis of Poissonian resetting (via the perspective of the inspection paradox) \cite{Pal2022}.

Standing at the positive time point $t_{0}$, also consider the inter-renewal period that `covers' the time point $t_{0}$, i.e.: the duration between the last renewal epoch before $t_{0}$, and the first renewal epoch after $t_{0}$. This inter-renewal period converges in law, as $t_{0}\rightarrow \infty $, to a limiting random variable that is termed the \emph{total lifetime} of the input $T$. The density function of the total lifetime is \cite{Cox}-\cite{Ros}:
\begin{equation}
f_{tot}\left( t\right) =\frac{1}{\mu }tf\left( t\right) \text{ ,}
\label{502}
\end{equation}
($t>0$). The total lifetime played a principal role in the mean-performance analysis of sharp restart (doing so via a socioeconomic perspective\footnote{To describe the socioeconomic interpretation of the total lifetime, consider a human society comprising members with positive wealth values. Sampling at random a society member, further consider $T$ to be the wealth of the randomly-sampled member. Now, sample at random a Dollar from the society's overall wealth, and set $T_{\$}$ to be the wealth of the society member to whom the randomly-sampled Dollar belongs. The random variable $T_{\$}$ is equal, in law, to the total lifetime of the input $T$. Namely, the density function of the random variable $T_{\$}$ is $f_{tot}(t)$.}) \cite{MP2SR}.

Two additional density functions correspond to the maximum and the minimum of $n$ independent and identically distributed copies $\left\{ T_{1},\cdots, T_{n}\right\} $ of the input $T$. The density function of the maximum is:
\begin{equation}
f_{\max }\left( t\right) =n F\left( t\right)^{n-1} f\left( t\right) \text{ ,}
\label{503}
\end{equation}
($t>0$); this density function follows, by differentiation, from the fact that the maximum's distribution function is $F(t)^{n}$ ($t>0$). The density function of the minimum is:
\begin{equation}
f_{\min }\left( t\right) =n \bar{F}\left( t\right)^{n-1} f\left( t\right) \text{ ,}
\label{504}
\end{equation}
($t>0$); this density function follows, by differentiation, from the fact that the minimum's survival function is $\bar{F}(t)^{n}$ ($t>0$).

\subsection{Relative entropy}

Relative entropy -- a.k.a. Kullback-Leibler divergence \cite{KL}-\cite{Kul} -- measures the `information distance' between two density functions that are defined over a common underlying set. Here the underlying set is the positive half-line ($t>0$). The Kullback-Leibler divergence of the density function $\phi \left(t\right)$ from the density function $\psi \left( t\right)$ is: 
\begin{equation}
D\left( \phi |\psi \right) =\int_{0}^{\infty } \ln \left[\frac{\phi \left( t\right) }{\psi \left( t\right) }\right] \phi \left( t\right)dt\text{ .}
\label{5000}
\end{equation}
The Gibbs inequality \cite{Fal}-\cite{Lie} asserts that the Kullback-Leibler divergence is non-negative, $D\left( \phi |\psi \right) \geq 0$, and that it vanishes if and only if the two density functions coincide:\footnote{To be mathematically precise: the equality $\phi \left( x\right) =\psi\left( x\right) $ should hold almost everywhere with respect to the Lebesgue measure (over the positive half-line).} $D\left( \phi |\psi \right)=0\Leftrightarrow \phi \left( x\right) =\psi \left( x\right) $. 

We point out that the Kullback-Leibler divergence is not a metric (in the sense of metric spaces). Indeed, the Kullback-Leibler divergence does not measure distance in a symmetric fashion: in general, $D\left( \phi |\psi \right) \neq D\left( \psi|\phi \right) $. Namely, the Kullback-Leibler distance of the density function $\phi \left(x\right) $ from the density function $\psi \left( x\right) $ need not be equal to the Kullback-Leibler distance of the density function $\psi \left( x\right) $ from the density function $\phi \left(x\right) $. Also, the Kullback-Leibler divergence does not (necessarily) exhibit the triangle inequality -- and this feature will play a principal role below.

The existence results to be presented below will use the Kullback-Leibler divergence in the following way: comparing the Kullback-Leibler distance $D(\phi | f_{res})$ to the sum of the Kullback-Leibler distances $D(\phi | f) + D(f | f_{res})$; where $\phi$ is one of the density functions $ \left\{ f_{tot},f_{\max},f_{\min} \right\}$. This comparison of Kullback-Leibler distances has a geometric interpretation.

In the space of density functions that are defined over the positive half-line, the distance $D(\phi | f_{res})$ manifests the length of the path $\phi \mapsto f_{res}$, i.e.: going directly from $\phi$ to $f_{res}$. And, the sum of the Kullback-Leibler distances $D(\phi | f) + D(f | f_{res})$ manifests the length of the path $\phi \mapsto f \mapsto f_{res}$, i.e.: going from $\phi$ to $f$, and then going from $f$ to $f_{res}$. As the Kullback-Leibler divergence does not necessarily exhibit the triangle inequality, the direct path $\phi \mapsto f_{res}$ can be either shorter or longer than the indirect path $\phi \mapsto f \mapsto f_{res}$ (Fig. \ref{KLFig}). The Kullback-Leibler terms $D(\phi | f)$ and $D(f | f_{res})$ -- whose sum is the length of the indirect path $\phi \mapsto f \mapsto f_{res}$ -- have profound entropy meanings which we shall now explain.

The Kullback-Leibler term $D(\phi | f)$ vanishes if and only if the density functions $\phi$ and $f$ coincide. As $\phi$ is one of the density functions $ \left\{ f_{tot},f_{\max},f_{\min} \right\}$, coincidence occurs if and only if the input $T$ is deterministic. Also, among all inputs with a given mean, the input with the minimal entropy is the deterministic input. Thus, in effect: the Kullback-Leibler term $D(\phi | f)$ manifests a ``distance from min-entropy''.

The Kullback-Leibler term $D(f | f_{res})$ vanishes if and only if the density functions $f$ and $f_{res}$ coincide. This coincidence occurs if and only if the input $T$ is exponentially distributed \cite{Cox}-\cite{Ros}. Also, among all inputs with a given mean, the input with the maximal entropy is the exponentially-distributed input. Thus, in effect: the Kullback-Leibler term $D(f | f_{res})$ manifests a ``distance from max-entropy''.

\subsection{Three existence criteria}

\begin{figure}[t!]
\centering
\includegraphics[width=8cm]{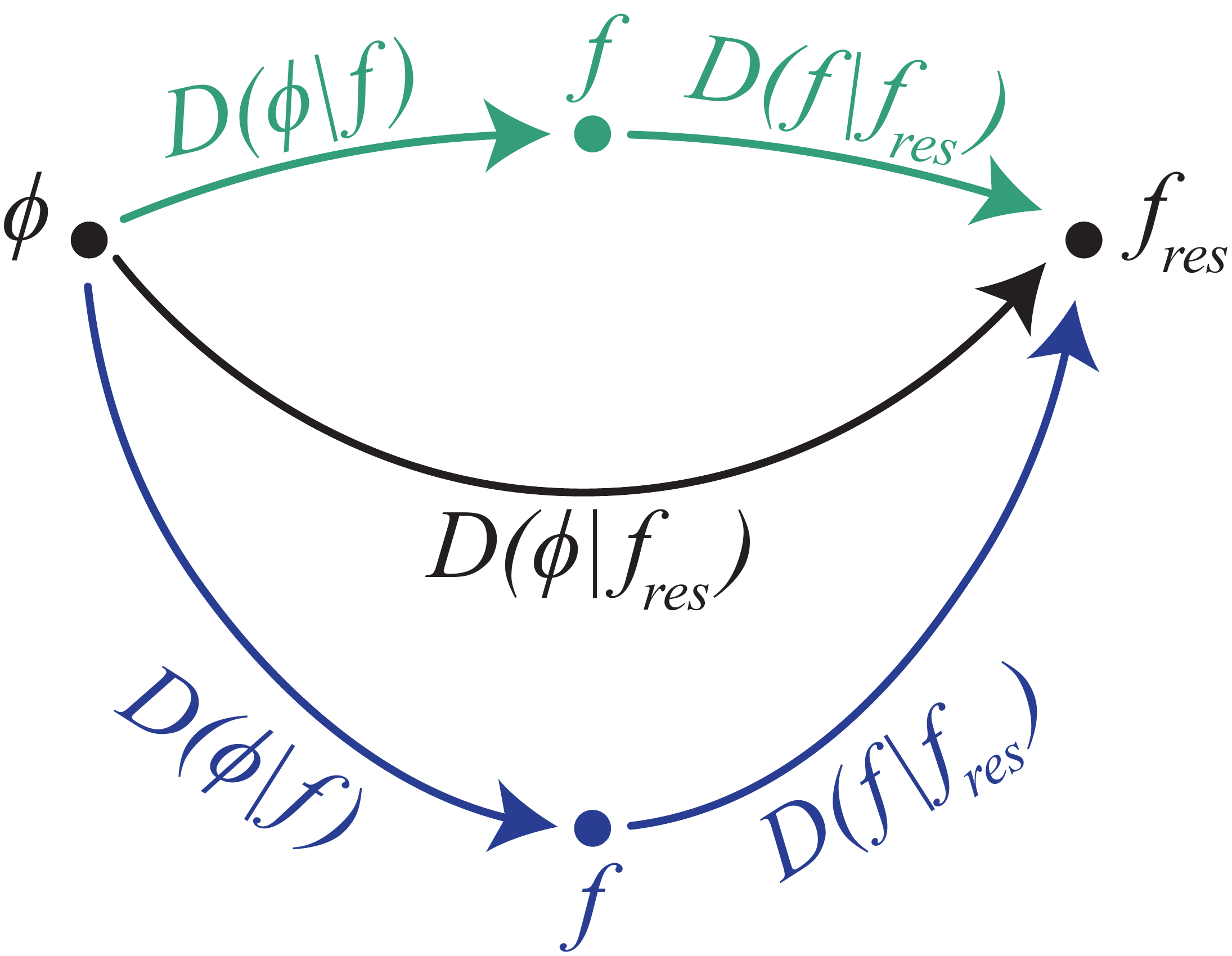}
\caption{Kullback-Leibler divergence. Distances measured by the Kullback-Leibler divergence do not necessarily obey the triangle inequality. Hence, in terms of the Kullback-Leibler divergence, the length of the direct path $\phi \mapsto f_{res}$ (in black) can be: either shorter (in turquoise), or longer (in blue), than the length of the indirect path $\phi \mapsto f \mapsto f_{res}$.}
\label{KLFig}
\end{figure}

Having introduced the four density functions $ \left\{f_{res},f_{tot},f_{\max},f_{\min} \right\}$, and having reviewed the notion of relative entropy, we are now all set to present the existence criteria. These criteria stem from a common integral:
\begin{equation}
I = \int_{0}^{\infty }\left[ E\left( \tau \right) -\eta \right] [F(\tau) w\left( \tau \right)] d\tau  \text{ ,}  \label{500}
\end{equation}
where $w(\tau)$ is a positive-valued weight function. Namely, the integral $I$ is the weighted average of $[E\left( \tau \right) -\eta]$ -- the difference between the output's entropy and the input's entropy -- where the weights are $F(\tau) w(\tau)$.

The existence criteria come in three pairs. Each pair emanates from Eq. (\ref{500}) via a specific weight function, and due to the following straightforward observations. If the integral is negative $I<0$ then the difference of entropies $[E\left( \tau \right) -\eta]$ must be negative for some $\tau$, and hence: there exist timers $\tau $ with which sharp restart decreases entropy. And, if the integral is positive $I>0$ then the difference of entropies $[E\left( \tau \right) -\eta]$ must be positive for some $\tau$, and hence: there exist timers $\tau $ with which sharp restart increases entropy.


Multiplying both sides of Eq. (\ref{301}) by the weight function $w(\tau)=F(\tau) \frac{1}{\mu}$, and then integrating over $\tau>0$ by parts, yields
\begin{equation}
I =D\left( f_{tot}|f_{res}\right) - [D\left(f_{tot}|f\right) + D\left( f|f_{res}\right)] \text{ .}  \label{510}
\end{equation}
In turn, Eq. (\ref{510}) implies the following pair of existence criteria.

\begin{enumerate}
\item[$\bullet $] If the direct path $ f_{tot} \mapsto f_{res}$ is shorter than the indirect path $ f_{tot} \mapsto f \mapsto f_{res}$ then there exist timers $\tau $ with which sharp restart decreases entropy.

\item[$\bullet $] If the direct path $ f_{tot} \mapsto f_{res}$ is longer than the indirect path $ f_{tot} \mapsto f \mapsto f_{res}$ then there exist timers $\tau $ with which sharp restart increases entropy.
\end{enumerate}

To demonstrate Eq. (\ref{510}) `in action', consider the following example: an input whose statistics are Pareto type I, with a finite mean. In this example the input's survival function is $\bar{F}(t)=t^{-p}$ ($t \geq 1$), where $p$ is a power that is larger than one ($p>1$). For this example, the density functions that take part in Eq. (\ref{510}) are: $f(t)=pt^{-p-1}$ ($t \geq 1$); $f_{tot}(t)=(p-1)t^{-p}$ ($t \geq 1$); and $f_{res}(t)=\frac{p-1}{p}t^{-p}$ ($t \geq 1$). A calculation involving these density functions implies that Eq. (\ref{510}) yields the value $I=-1/[p(p-1)]$, and hence: there exist timers $\tau$ with which sharp restart decreases entropy. This conclusion is in accord with the results following Eq. (\ref{422}). Indeed, in the Pareto type I example $H\left( \infty \right) =0$, and thus Eq. (\ref{422}) implies that: sharp restart with sufficiently large timers $\tau $ decreases entropy.

Multiplying both sides of Eq. (\ref{301}) by the weight function $w(\tau)=n(n-1)F(\tau)^{n-1}f(\tau)$, and then integrating over $\tau>0$ by parts, yields
\begin{equation}
I =D\left( f_{\max }|f_{res}\right) - [D\left( f_{\max}|f\right) + D\left( f|f_{res}\right)] \text{ .}  
\label{520}
\end{equation}
In turn, Eq. (\ref{520}) implies the following pair of existence
criteria.

\begin{enumerate}
\item[$\bullet $] If the direct path $ f_{\max} \mapsto f_{res}$ is shorter than the indirect path $ f_{\max} \mapsto f \mapsto f_{res}$ then there exist timers $\tau $ with which sharp restart decreases entropy.

\item[$\bullet $] If the direct path $ f_{\max} \mapsto f_{res}$ is longer than the indirect path $ f_{\max} \mapsto f \mapsto f_{res}$ then there exist timers $\tau $ with which sharp restart increases entropy.
\end{enumerate}

Multiplying both sides of Eq. (\ref{302}) by the weight function $w(\tau)=n(n-1)F(\tau)\bar{F}(\tau)^{n-2}f(\tau)$, and then integrating over $\tau>0$ by parts, yields
\begin{equation}
I = [D\left( f_{\min}|f\right) + D\left( f|f_{res}\right)] - D\left( f_{\min }|f_{res}\right) \text{ .}  
\label{530}
\end{equation}
In turn, Eq. (\ref{530}) implies the following pair of existence
criteria.

\begin{enumerate}
\item[$\bullet $] If the direct path $ f_{\min} \mapsto f_{res}$ is longer than the indirect path $ f_{\min} \mapsto f \mapsto f_{res}$ then there exist timers $\tau $ with which sharp restart decreases entropy.

\item[$\bullet $] If the direct path $ f_{\min} \mapsto f_{res}$ is shorter than the indirect path $ f_{\min} \mapsto f \mapsto f_{res}$ then there exist timers $\tau $ with which sharp restart increases entropy.
\end{enumerate}

The proofs of Eqs. (\ref{510})-(\ref{530}) are detailed in the Methods. To conclude this section, we address the case of Exponentially-distributed inputs. Recall that, as established in section \ref{3} above, sharp restart has no effect on entropy -- i.e. $E( \tau)=\eta$ for all timers $\tau$ -- if and only if the input $T$ is Exponentially-distributed.

As noted above, the density functions $f$ and $f_{res}$ coincide if and only if the input $T$ is exponentially distributed \cite{Cox}-\cite{Ros}. So, on the one hand, if the input is Exponentially-distributed then: setting $E( \tau)=\eta$ ($\tau>0$) in Eq. (\ref{500}) yields a zero integral, $I=0$, for any weight function $w(\tau)$. On the other hand, if the input is Exponentially-distributed then: as $f=f_{res}$, the direct path $ \phi \mapsto f_{res}$ has the same length as the indirect path $ \phi \mapsto f \mapsto f_{res}$ -- i.e. $D(\phi | f_{res})=D( \phi | f)+D(f | f_{res})$ -- for any density function $\phi$ (which is defined over the positive half-line).

Thus, the existence criteria of this section are in accord with the aforementioned result of section \ref{3}. Indeed, if the input is Exponentially-distributed then this section's criteria imply neither the existence of timers with which sharp restart decreases entropy, nor the existence of timers with which sharp restart increases entropy.

\section{\label{6}Summary}

This paper presented a comprehensive entropy-based stochasticity analysis of sharp restart. When viewed from an algorithmic perspective, sharp restart can be described as a non-linear map that: receives an input $T$ which is a positive-valued random variable; and -- using a positive timer parameter $\tau $ -- produces an output $T_{R}$ which is also a positive-valued random variable. Specifically, the input $T$ manifest the random time required to accomplish a task of interest. As long as the task is not accomplished, the algorithm restarts the task every $\tau $ time-units. So, \emph{under sharp restart}, the random time required to accomplish the task is the output $T_{R}$.

The stochasticity analysis compared the output's entropy $E\left( \tau \right) $ (which is a function of the timer parameter $\tau $) to the input's entropy $\eta $. The two principal analytic tools employed were the \emph{hazard rate}\ of reliability engineering, and the \emph{Kullback-Leibler divergence} of information theory. The results established provide an `entropy roadmap' for sharp restart: seven pairs of universal criteria that determine if the application of sharp restart decreases the entropy, $E\left( \tau \right) <\eta $, or if it increases the entropy, $E\left( \tau \right) >\eta $. These pairs of criteria are summarized in Table 1.

The pairs of criteria appearing in rows \textbf{I}-\textbf{IV} of Table 1 all involve the rate $r_{exp}=e/\exp(\eta)$, which manifests an `Exponential benchmark' for the input. Specifically, $r_{exp}$ is the constant hazard rate of an Exponential distribution whose entropy is equal to the input's entropy $\eta$. The criteria of rows \textbf{I}-\textbf{IV} compare the input's hazard rate to the Exponential benchmark $r_{exp}$. Thus, this Exponential benchmark assumes a key role in determining the effect of sharp restart on entropy. The pairs of criteria appearing in rows \textbf{V}-\textbf{VII} of Table 1 address the very existence of timers with which sharp restart decreases or increases entropy.

As noted in section \ref{3}, the exponentiation of the input's entropy -- the quantity $\exp(\eta)$ -- is the input's \emph{perplexity} \cite{JMB}. The input's Boltzmann-Gibbs-Shannon entropy $\eta$ and perplexity $\exp(\eta)$ are special cases of, respectively, the input's \emph{Renyi entropy} \cite{Ren}-\cite{Zyc} and \emph{diversity} \cite{Hil}-\cite{LL}. Elevating from the Boltzmann-Gibbs-Shannon entropy to the Renyi entropy, the second part of this duo will present a comprehensive diversity-based stochasticity analysis of sharp restart.

\bigskip

\textbf{Acknowledgments}. The authors thank Shira Yovel for help in producing Figure 1. Shlomi Reuveni acknowledges support from the Israel Science Foundation (grant No. 394/19). This project has received funding from the European Research Council (ERC) under the European Union’s Horizon 2020 research and innovation program (Grant agreement No. 947731).  \\

\newpage

\begin{center}
\ \ 

{\LARGE Table 1 }

\ \ \ \ \ \ 

\begin{tabular}{|l|l|l|l|l|}
\hline
& $%
\begin{array}{c}
\text{\textbf{\ }} \\ 
\text{\textbf{Timer}} \\ 
\text{\textbf{\ }}%
\end{array}%
$ & $%
\begin{array}{c}
\text{\textbf{\ }} \\ 
\text{\textbf{Parameter}} \\ 
\text{\textbf{\ }}%
\end{array}%
$ & $%
\begin{array}{c}
\text{\textbf{\ }} \\ 
\text{\textbf{Decrease}} \\ 
\text{\textbf{\ }}%
\end{array}%
$ & $%
\begin{array}{c}
\text{\textbf{\ }} \\ 
\text{\textbf{Increase}} \\ 
\text{\textbf{\ }}%
\end{array}%
$ \\ \hline
\textbf{I} & $%
\begin{array}{c}
\text{ } \\ 
\text{General} \\ 
\text{ }%
\end{array}%
$ & $0<\tau <\infty $ & $\int_{0}^{\tau }\ln \left[ \frac{H\left( t\right) }{r_{\exp}}\right] f\left( t\right) dt>0$ & $\int_{0}^{\tau }\ln \left[ \frac{H\left(
t\right) }{r_{\exp}}\right] f\left( t\right) dt<0$ \\ \hline
\textbf{II} & $%
\begin{array}{c}
\text{ } \\ 
\text{General} \\ 
\text{ }%
\end{array}%
$ & $0<\tau <\infty $ & $\int_{\tau }^{\infty }\ln \left[ \frac{H\left(
t\right) }{r_{\exp}}\right] f\left( t\right) dt<0$ & $\int_{\tau }^{\infty }\ln \left[ \frac{H\left( t\right) }{r_{\exp}}\right] f\left( t\right) dt>0$ \\ \hline
\textbf{III} & $%
\begin{array}{c}
\text{ } \\ 
\text{Fast} \\ 
\text{ }%
\end{array}%
$ & $0<\tau <\tau _{\ast }$ & $H\left( 0\right) >r_{\exp}$ & $H\left( 0\right) <r_{\exp}$
\\ \hline
\textbf{IV} & $%
\begin{array}{c}
\text{ } \\ 
\text{Slow} \\ 
\text{ }%
\end{array}%
$ & $\tau ^{\ast }<\tau <\infty $ & $H\left( \infty \right) <r_{\exp}$ & $H\left(
\infty \right) >r_{\exp}$ \\ \hline
\textbf{V} & $%
\begin{array}{c}
\text{ } \\ 
\text{Existence} \\ 
\text{ }%
\end{array}%
$ & ------------ & $%
\begin{array}{c}
\text{ } \\ 
D\left( f_{tot}|f_{res}\right) < \\ 
\text{ } \\ 
D\left( f_{tot}|f\right) +D\left( f|f_{res}\right)  \\ 
\text{ }%
\end{array}%
$ & $%
\begin{array}{c}
\text{ } \\ 
D\left( f_{tot}|f_{res}\right) > \\ 
\text{ } \\ 
D\left( f_{tot}|f\right) +D\left( f|f_{res}\right)  \\ 
\text{ }%
\end{array}%
$ \\ \hline
\textbf{VI} & $%
\begin{array}{c}
\text{ } \\ 
\text{Existence} \\ 
\text{ }%
\end{array}%
$ & ------------ & $%
\begin{array}{c}
\text{ } \\ 
D\left( f_{\max }|f_{res}\right) < \\ 
\text{ } \\ 
D\left( f_{\max }|f\right) +D\left( f|f_{res}\right)  \\ 
\text{ }%
\end{array}%
$ & $%
\begin{array}{c}
\text{ } \\ 
D\left( f_{\max }|f_{res}\right) > \\ 
\text{ } \\ 
D\left( f_{\max }|f\right) +D\left( f|f_{res}\right)  \\ 
\text{ }%
\end{array}%
$ \\ \hline
\textbf{VII} & $%
\begin{array}{c}
\text{ } \\ 
\text{Existence} \\ 
\text{ }%
\end{array}%
$ & ------------ & $%
\begin{array}{c}
\text{ } \\ 
D\left( f_{\min }|f_{res}\right) > \\ 
\text{ } \\ 
D\left( f_{\min }|f\right) +D\left( f|f_{res}\right)  \\ 
\text{ }%
\end{array}%
$ & $%
\begin{array}{c}
\text{ } \\ 
D\left( f_{\min }|f_{res}\right) < \\ 
\text{ } \\ 
D\left( f_{\min }|f\right) +D\left( f|f_{res}\right)  \\ 
\text{ }%
\end{array}%
$ \\ \hline
\end{tabular}
\end{center}

\bigskip\

\textbf{Table 1}: Seven pairs of universal criteria that determine the effect of sharp restart on entropy. The columns specify the key features of each pair of criteria: to which timer parameters $\tau$ the criteria apply, and when does the application of sharp restart decrease/increase the entropy (of the output, with respect to that of the input). Rows \textbf{I} and \textbf{II} -- criteria for general timers (section \ref{3}), where: $f(t)$ and $H(t)$ are, respectively, the input's density and hazard functions; the rate $r_{\exp}=e/\exp(\eta)$ is the height of the flat hazard function that characterizes an Exponential distribution whose entropy equal to the input's entropy $\eta$. Rows \textbf{III} and \textbf{IV} -- criteria for fast and slow timers (section \ref{4}), where: $H(0)$ and $H(\infty)$ are the limit values of the input's hazard function; the threshold $\tau_{*}$ is the upper bound of the range of fast timers (see Eq. (\ref{413})); the threshold $\tau^{*}$ is the lower bound of the range of slow timers (see Eq. (\ref{423})). Rows \textbf{V}-\textbf{VII} -- existence criteria (section \ref{5}): criteria that determine the very existence of timers with which sharp restart decreases/increases the entropy. The existence criteria employ the Kullback-Leibler divergence $D( \cdot | \cdot )$ to measure the relative entropies between the following density functions: $f$ of the input; $f_{res}$ of the input's residual lifetime (see Eq. (\ref{501}); $f_{tot}$ of the input's total lifetime (see Eq. (\ref{502}); and $f_{max}$ and $f_{min}$ of the maximum and minimum, respectively, of $n$ IID copies of the input (see Eqs. (\ref{503}) and (\ref{504})).

\bigskip

\newpage

\section{\label{7}Methods}

\subsection{Derivation of Eq. (\protect\ref{202})}

Set $p=F\left( \tau \right) $ and $q=\bar{F}\left( \tau \right) $. Note that
\begin{equation}
\int_{0}^{\tau }f\left( u\right) du=F\left( \tau \right) =p\text{ .}
\label{A101}
\end{equation}
Using the periodic parameterization of the time axis $t=\tau n+u$ (in which $n=0,1,2,\cdots $ and $0\leq u<\tau $), Eq. (\ref{201}) implies that 
\begin{equation}
f_{R}\left( \tau n+u\right) =q^{n}f\left( u\right) \text{ .}  \label{A102}
\end{equation}
Eqs. (\ref{A101}) and (\ref{A102}) imply that:
\begin{equation}
\left. 
\begin{array}{l}
\int_{0}^{\infty }f_{R}\left( t\right) \ln \left[ f_{R}\left( t\right)\right] dt=\sum_{n=0}^{\infty }\int_{\tau n}^{\tau n+\tau }f_{R}\left(t\right) \ln \left[ f_{R}\left( t\right) \right] dt \\ 
\text{ } \\ 
=\sum_{n=0}^{\infty }\int_{0}^{\tau }f_{R}\left( \tau n+u\right) \ln \left[f_{R}\left( \tau n+u\right) \right] du \\ 
\text{ } \\ 
=\sum_{n=0}^{\infty }\int_{0}^{\tau }\left[ q^{n}f\left( u\right) \right] \ln \left[ q^{n}f\left( u\right) \right] du \\ 
\text{ } \\ 
=\sum_{n=0}^{\infty }q^{n}\int_{0}^{\tau }f\left( u\right) \left\{ n\ln \left( q\right) +\ln \left[ f\left( u\right) \right] \right\} du \\ 
\text{ } \\ 
=\sum_{n=0}^{\infty }q^{n}\left\{ n\ln \left( q\right) \int_{0}^{\tau}f\left( u\right) du+\int_{0}^{\tau }f\left( u\right) \ln \left[ f\left(u\right) \right] du\right\} \\ 
\text{ } \\ 
=\left[ \sum_{n=0}^{\infty }nq^{n}p\right] \ln \left( q\right) +\frac{1}{p} \left[ \sum_{n=0}^{\infty }q^{n}p\right] \left[ \int_{0}^{\tau }f\left(u\right) \ln \left[ f\left( u\right) \right] du\right] \text{ .}
\end{array}
\right.  \label{A103}
\end{equation}

Consider a Geometric random variable $N$, over the non-negative integers, with success probability $p$. The probability distribution of this random variable is given by $\Pr \left( N=n\right) =q^{n}p$ ($n=0,1,2,\cdots $), and hence
\begin{equation}
\sum_{n=0}^{\infty }q^{n}p=\sum_{n=0}^{\infty }\Pr \left( N=n\right) =1\text{
.}  \label{A104}
\end{equation}
Moreover, the mean of the random variable $N$ is 
\begin{equation}
\sum_{n=0}^{\infty }nq^{n}p=\mathbf{E}\left[ N\right] =\frac{q}{p}\text{ .}
\label{A105}
\end{equation}
Substituting Eqs. (\ref{A104}) and (\ref{A105}) into the bottom line of Eq. (\ref{A103}) yields
\begin{equation}
\left. 
\begin{array}{l}
\int_{0}^{\infty }f_{R}\left( t\right) \ln \left[ f_{R}\left( t\right) 
\right] dt \\ 
\text{ } \\ 
=\frac{q}{p}\ln \left( q\right) +\frac{1}{p}\int_{0}^{\tau }f\left( u\right) \ln \left[ f\left( u\right) \right] du \\ 
\text{ } \\ 
=\frac{1}{F\left( \tau \right) }\left\{\bar{F}\left( \tau \right) \ln \left[
\bar{F}\left( \tau \right) \right] +\int_{0}^{\tau }f\left( t\right) \ln 
\left[ f\left( t\right) \right] dt\right\} \text{ .}
\end{array}
\right.  \label{A106}
\end{equation}
In turn, Eq. (\ref{A106}) yields Eq. (\ref{202}).

\subsection{Derivation of Eq. (\protect\ref{203})}

Eq. (\ref{202}) implies that
\begin{equation}
\left. 
\begin{array}{l}
\left[ E\left( \tau \right) -\eta \right] F\left( \tau \right) \\ 
\text{ } \\ 
=-\bar{F}\left( \tau \right) \ln \left[ \bar{F}\left( \tau \right) \right]
-\int_{0}^{\tau }f\left( t\right) \ln \left[ f\left( t\right) \right]
dt-\eta F\left( \tau \right) \\ 
\text{ } \\ 
=-\bar{F}\left( \tau \right) \ln \left[ \bar{F}\left( \tau \right) \right]
-\int_{0}^{\tau }f\left( t\right) \ln \left[ f\left( t\right) \right]
dt-\eta \left[ 1-\bar{F}\left( \tau \right) \right] \\ 
\text{ } \\ 
=-\bar{F}\left( \tau \right) \ln \left[ \bar{F}\left( \tau \right) \right]
-\left\{ \eta +\int_{0}^{\tau }f\left( t\right) \ln \left[ f\left( t\right) 
\right] dt\right\} +\eta \bar{F}\left( \tau \right) \text{ .}
\end{array}
\right.  \label{A111}
\end{equation}
Note that
\begin{equation}
\left. 
\begin{array}{l}
\eta +\int_{0}^{\tau }f\left( t\right) \ln \left[ f\left( t\right) \right] dt
\\ 
\text{ } \\ 
=-\int_{0}^{\infty }f\left( t\right) \ln \left[ f\left( t\right) \right]
dt+\int_{0}^{\tau }f\left( t\right) \ln \left[ f\left( t\right) \right] dt
\\ 
\text{ } \\ 
=-\int_{\tau }^{\infty }f\left( t\right) \ln \left[ f\left( t\right) \right]
dt\text{ .}
\end{array}
\right.  \label{A112}
\end{equation}
Substituting Eq. (\ref{A112}) into the bottom line of Eq. (\ref{A111}) yields
\begin{equation}
\left[ E\left( \tau \right) -\eta \right] F\left( \tau \right) =\bar{F}
\left( \tau \right) \left\{ \eta -\ln \left[ \bar{F}\left( \tau \right) 
\right] +\frac{1}{\bar{F}\left( \tau \right) }\int_{\tau }^{\infty }f\left(
t\right) \ln \left[ f\left( t\right) \right] dt\right\} \text{ .}
\label{A113}
\end{equation}
In turn, Eq. (\ref{A113}) yields Eq. (\ref{203}).

\subsection{Derivation of Eqs. (\protect\ref{3001}) and (\protect\ref{3002})}

As $H\left( t\right) =f\left( t\right) /\bar{F}\left( t\right) $, note that%
\begin{equation}
\left. 
\begin{array}{l}
\int_{0}^{\tau }f\left( t\right) \ln \left[ f\left( t\right) \right]
dt=\int_{0}^{\tau }f\left( t\right) \ln \left[ \bar{F}\left( t\right)
H\left( t\right) \right] dt \\ 
\text{ } \\ 
=\int_{0}^{\tau }\ln \left[ \bar{F}\left( t\right) \right] f\left( t\right)
dt+\int_{0}^{\tau }\ln \left[ H\left( t\right) \right] f\left( t\right) dt%
\text{ .}%
\end{array}%
\right.  \label{B101}
\end{equation}%
Using the change-of variables $t\mapsto u=\bar{F}\left( t\right) $, further
note that%
\begin{equation}
\left. 
\begin{array}{l}
\int_{0}^{\tau }\ln \left[ \bar{F}\left( t\right) \right] f\left( t\right)
dt=\int_{\bar{F}\left( \tau \right) }^{1}\ln \left( u\right) du \\ 
\text{ } \\ 
=\left. \overset{\text{ }}{\left[ u\ln \left( u\right) -u\right] }%
\right\vert _{\bar{F}\left( \tau \right) }^{1}=-1-\bar{F}\left( \tau \right)
\ln \left[ \bar{F}\left( \tau \right) \right] +\bar{F}\left( \tau \right) \\ 
\text{ } \\ 
=-F\left( \tau \right) -\bar{F}\left( \tau \right) \ln \left[ \bar{F}\left(
\tau \right) \right] \text{ .}%
\end{array}%
\right.  \label{B102}
\end{equation}%
Substituting Eq. (\ref{B102}) into Eq. (\ref{B101}) yields%
\begin{equation}
\left. 
\begin{array}{l}
\int_{0}^{\tau }f\left( t\right) \ln \left[ f\left( t\right) \right] dt \\ 
\text{ } \\ 
=\left\{ -F\left( \tau \right) -\bar{F}\left( \tau \right) \ln \left[ \bar{F}%
\left( \tau \right) \right] \right\} +\int_{0}^{\tau }\ln \left[ H\left(
t\right) \right] f\left( t\right) dt\text{ .}%
\end{array}%
\right.  \label{B103}
\end{equation}%
In turn, Eq. (\ref{B103}) implies that 
\begin{equation}
\left. 
\begin{array}{l}
\bar{F}\left( \tau \right) \ln \left[ \bar{F}\left( \tau \right) \right]
+\int_{0}^{\tau }f\left( t\right) \ln \left[ f\left( t\right) \right] dt \\ 
\text{ } \\ 
=-F\left( \tau \right) +\int_{0}^{\tau }\ln \left[ H\left( t\right) \right]
f\left( t\right) dt \\ 
\text{ } \\ 
=-\left\{ \int_{0}^{\tau }f\left( t\right) dt-\int_{0}^{\tau }\ln \left[
H\left( t\right) \right] f\left( t\right) dt\right\} \\ 
\text{ } \\ 
=-\int_{0}^{\tau }\left\{ 1-\ln \left[ H\left( t\right) \right] \right\}
f\left( t\right) dt \\ 
\text{ } \\ 
=-\int_{0}^{\tau }\ln \left[ \frac{e}{H\left( t\right) }\right] f\left(
t\right) dt\text{ .}%
\end{array}%
\right.  \label{B104}
\end{equation}%
Finally, substituting Eq. (\ref{B104}) into Eq. (\ref{202}) yields%
\begin{equation}
\left. 
\begin{array}{l}
E\left( \tau \right) =\frac{-1}{F\left( \tau \right) }\left\{ \bar{F}\left(
\tau \right) \ln \left[ \bar{F}\left( \tau \right) \right] +\int_{0}^{\tau
}f\left( t\right) \ln \left[ f\left( t\right) \right] dt\right\} \\ 
\text{ } \\ 
=\frac{1}{F\left( \tau \right) }\int_{0}^{\tau }\ln \left[ \frac{e}{H\left(
t\right) }\right] f\left( t\right) dt=\int_{0}^{\tau }\ln \left[ \frac{e}{%
H\left( t\right) }\right] \left[ \frac{f\left( t\right) }{F\left( \tau
\right) }\right] dt\text{ .}%
\end{array}%
\right.  \label{B105}
\end{equation}%
Eq. (\ref{B105}) proves Eq. (\ref{3002}). And, in particular, setting $\tau
=\infty $ yields Eq. (\ref{3001}):%
\begin{equation}
\eta =\int_{0}^{\infty }\ln \left[ \frac{e}{H\left( t\right) }\right]
f\left( t\right) dt\text{ .}  \label{B106}
\end{equation}

\subsection{Derivation of Eqs. (\protect\ref{301}) and (\protect\ref{302})}

Eq. (\ref{B105}) implies Eq. (\ref{301}):%
\begin{equation}
\left. 
\begin{array}{l}
E\left( \tau \right) -\eta =\frac{1}{F\left( \tau \right) }\int_{0}^{\tau
}\ln \left[ \frac{e}{H\left( t\right) }\right] f\left( t\right) dt-\eta \\ 
\text{ } \\ 
=\frac{1}{F\left( \tau \right) }\left\{ \int_{0}^{\tau }\ln \left[ \frac{e}{%
H\left( t\right) }\right] f\left( t\right) dt-\eta \int_{0}^{\tau }f\left(
t\right) dt\right\} \\ 
\text{ } \\ 
=\frac{1}{F\left( \tau \right) }\int_{0}^{\tau }\left\{ \ln \left[ \frac{e}{%
H\left( t\right) }\right] -\eta \right\} f\left( t\right) dt \\ 
\text{ } \\ 
=\frac{1}{F\left( \tau \right) }\int_{0}^{\tau }\ln \left[ \frac{e^{1-\eta }%
}{H\left( t\right) }\right] f\left( t\right) dt \\ 
\text{ } \\ 
=\frac{1}{F\left( \tau \right) }\int_{0}^{\tau }\ln \left[ \frac{r_{\exp}}{H\left(
t\right) }\right] f\left( t\right) dt%
\end{array}%
\right.  \label{B111}
\end{equation}%
(recall that $r_{\exp}=\exp \left( 1-\eta \right) $). Setting $\tau =\infty $ in
Eq. (\ref{B111}) yields 
\begin{equation}
\left. 
\begin{array}{l}
0=\int_{0}^{\infty }\ln \left[ \frac{r_{\exp}}{H\left( t\right) }\right] f\left(
t\right) dt \\ 
\text{ } \\ 
=\int_{0}^{\tau }\ln \left[ \frac{r_{\exp}}{H\left( t\right) }\right] f\left(
t\right) dt+\int_{\tau }^{\infty }\ln \left[ \frac{r_{\exp}}{H\left( t\right) }%
\right] f\left( t\right) dt\text{ .}%
\end{array}%
\right.  \label{B112}
\end{equation}%
In turn, Eq. (\ref{B112}) implies that%
\begin{equation}
\int_{0}^{\tau }\ln \left[ \frac{r_{\exp}}{H\left( t\right) }\right] f\left(
t\right) dt=\int_{\tau }^{\infty }\ln \left[ \frac{H\left( t\right) }{r_{\exp}}%
\right] f\left( t\right) dt\text{ .}  \label{B113}
\end{equation}%
Substituting Eq. (\ref{B113}) into Eq. (\ref{B111}) yields Eq. (\ref{302}): 
\begin{equation}
E\left( \tau \right) -\eta =\frac{1}{F\left( \tau \right) }\int_{\tau
}^{\infty }\ln \left[ \frac{H\left( t\right) }{r_{\exp}}\right] f\left( t\right) dt%
\text{ .}  \label{B114}
\end{equation}

\subsection{Kullback-Leibler calculations}

Consider three density functions that are defined over the positive
half-line ($t>0$): $\phi \left( t\right) $ and $\psi \left( t\right) $ (as
in Eq. (\ref{5000})), and $\varphi \left( t\right) $. Note that:
\begin{equation}
\left. 
\begin{array}{l}
\int_{0}^{\infty }\varphi \left( t\right) \ln \left[ \frac{\phi \left(
t\right) }{\psi \left( t\right) }\right] dt=\int_{0}^{\infty }\varphi \left(
t\right) \ln \left[ \frac{\phi \left( t\right) }{\varphi \left( t\right) }%
\frac{\varphi \left( t\right) }{\psi \left( t\right) }\right] dt \\ 
\text{ } \\ 
=\int_{0}^{\infty }\varphi \left( t\right) \left\{ \ln \left[ \frac{\phi
\left( t\right) }{\varphi \left( t\right) }\right] +\ln \left[ \frac{\varphi
\left( t\right) }{\psi \left( t\right) }\right] \right\} dt \\ 
\text{ } \\ 
=\int_{0}^{\infty }\varphi \left( t\right) \left\{ -\ln \left[ \frac{\varphi
\left( t\right) }{\phi \left( t\right) }\right] +\ln \left[ \frac{\varphi
\left( t\right) }{\psi \left( t\right) }\right] \right\} dt \\ 
\text{ } \\ 
=\int_{0}^{\infty }\varphi \left( t\right) \ln \left[ \frac{\varphi \left(
t\right) }{\psi \left( t\right) }\right] dt-\int_{0}^{\infty }\varphi \left(
t\right) \ln \left[ \frac{\varphi \left( t\right) }{\phi \left( t\right) }%
\right] dt \\ 
\text{ } \\ 
=D\left( \varphi |\psi \right) -D\left( \varphi |\phi \right) \text{ .}%
\end{array}%
\right.  \label{B120}
\end{equation}%
In transition to the bottom line of Eq. (\ref{B120}) we used the definition
of the Kullback-Leibler divergence.

We now turn to calculate the Kullback-Leibler divergence of the input's
density function $f\left( t\right) $ from the density function $%
f_{res}\left( t\right) =\frac{1}{\mu }\bar{F}\left( t\right) $ (which is the
density function of the input's residual lifetime). Note that: 
\begin{equation}
\left. 
\begin{array}{l}
D\left( f|f_{res}\right) =\int_{0}^{\infty }f\left( t\right) \ln \left[ 
\frac{f\left( t\right) }{f_{res}\left( t\right) }\right] dt \\ 
\text{ } \\ 
=\int_{0}^{\infty }f\left( t\right) \left\{ \ln \left[ f\left( t\right) %
\right] -\ln \left[ f_{res}\left( t\right) \right] \right\} dt \\ 
\text{ } \\ 
=\int_{0}^{\infty }f\left( t\right) \ln \left[ f\left( t\right) \right]
dt-\int_{0}^{\infty }f\left( t\right) \ln \left[ f_{res}\left( t\right) %
\right] dt \\ 
\text{ } \\ 
=-\eta -\int_{0}^{\infty }f\left( t\right) \ln \left[ f_{res}\left( t\right) %
\right] dt\text{ .}%
\end{array}%
\right.  \label{A140}
\end{equation}%
Also note that:%
\begin{equation}
\left. 
\begin{array}{l}
\int_{0}^{\infty }f\left( t\right) \ln \left[ f_{res}\left( t\right) \right]
dt=\int_{0}^{\infty }f\left( t\right) \ln \left[ \frac{1}{\mu }\bar{F}\left(
t\right) \right] dt \\ 
\text{ } \\ 
=\int_{0}^{\infty }f\left( t\right) \left\{ \ln \left[ \bar{F}\left(
t\right) \right] -\ln \left( \mu \right) \right\} dt \\ 
\text{ } \\ 
=\int_{0}^{\infty }f\left( t\right) \ln \left[ \bar{F}\left( t\right) \right]
dt-\ln \left( \mu \right) \int_{0}^{\infty }f\left( t\right) dt \\ 
\text{ } \\ 
=\int_{0}^{\infty }\ln \left[ \bar{F}\left( t\right) \right] f\left(
t\right) dt-\ln \left( \mu \right) \text{ .}%
\end{array}%
\right.  \label{A141}
\end{equation}%
Setting $\tau =\infty $ in Eq. (\ref{B102}) (and using the fact that $%
\lim_{u\rightarrow 0}\left[ u\ln \left( u\right) \right] =0$) implies that 
\begin{equation}
\int_{0}^{\infty }\ln \left[ \bar{F}\left( t\right) \right] f\left( t\right)
dt=-1\text{ .}  \label{A142}
\end{equation}%
Substituting Eq. (\ref{A142}) into Eq. (\ref{A141}) yields%
\begin{equation}
\int_{0}^{\infty }f\left( t\right) \ln \left[ f_{res}\left( t\right) \right]
dt=-1-\ln \left( \mu \right) \text{ .}  \label{A143}
\end{equation}%
In turn, substituting Eq. (\ref{A143}) into Eq. (\ref{A140}) (and recalling
that $r_{\exp}=\exp \left( 1-\eta \right) $) yields 
\begin{equation}
\left. 
\begin{array}{l}
D\left( f|f_{res}\right) =-\eta -\left[ -1-\ln \left( \mu \right) \right] \\ 
\text{ } \\ 
=\left( 1-\eta \right) +\ln \left( \mu \right) =\ln \left( r_{\exp}\right) +\ln
\left( \mu \right) \\ 
\text{ } \\ 
=\ln \left( \mu r_{\exp} \right) \text{ .}%
\end{array}%
\right.  \label{A144}
\end{equation}

\subsection{Derivation of Eqs. (\protect\ref{510}) and (\protect\ref{520})}

Let $w\left( \tau \right) $ ($\tau >0$) be a positive-valued `weight
function', set $W\left( t\right) =\int_{0}^{t}w\left( \tau \right) d\tau $ ($%
t>0$), and further set $g\left( t\right) =W\left( t\right) f\left( t\right) $
($t>0$). Multiplying both sides of Eq. (\ref{302}) by $F\left( \tau \right)
w\left( \tau \right) $, and integrating over the positive half-line, yields%
\begin{equation}
\left. 
\begin{array}{l}
\int_{0}^{\infty }\left[ E\left( \tau \right) -\eta \right] F\left( \tau
\right) w\left( \tau \right) d\tau \\ 
\text{ } \\ 
=\int_{0}^{\infty }\left\{ \int_{\tau }^{\infty }\ln \left[ \frac{H\left(
t\right) }{r_{\exp}}\right] f\left( t\right) dt\right\} w\left( \tau \right) d\tau
\\ 
\text{ } \\ 
=\int_{0}^{\infty }\ln \left[ \frac{H\left( t\right) }{r_{\exp}}\right] f\left(
t\right) \left\{ \int_{0}^{t}w\left( \tau \right) d\tau \right\} dt \\ 
\text{ } \\ 
=\int_{0}^{\infty }\ln \left[ \frac{H\left( t\right) }{r_{\exp}}\right] f\left(
t\right) W\left( t\right) dt \\ 
\text{ } \\ 
=\int_{0}^{\infty }\ln \left[ \frac{H\left( t\right) }{r_{\exp}}\right] g\left(
t\right) dt\text{ .}%
\end{array}%
\right.  \label{B131}
\end{equation}%
As $H\left( t\right) =f\left( t\right) /\bar{F}\left( t\right) $, and as $%
f_{res}\left( t\right) =\frac{1}{\mu }\bar{F}\left( t\right) $, note that%
\begin{equation}
\frac{H\left( t\right) }{r_{\exp}}=\frac{f\left( t\right) }{r_{\exp}\bar{F}\left( t\right) 
}=\frac{f\left( t\right) }{f_{res}\left( t\right) }\cdot \frac{1}{r_{\exp}\mu }%
\text{ ,}  \label{B132}
\end{equation}%
and hence%
\begin{equation}
\ln \left[ \frac{H\left( t\right) }{r_{\exp}}\right] =\ln \left[ \frac{f\left(
t\right) }{f_{res}\left( t\right) }\right] -\ln \left( r_{\exp}\mu \right) \text{ .}
\label{B133}
\end{equation}%
In turn, if $g\left( t\right) $ is a density function then%
\begin{equation}
\left. 
\begin{array}{l}
\int_{0}^{\infty }\ln \left[ \frac{H\left( t\right) }{r_{\exp}}\right] g\left(
t\right) dt \\ 
\text{ } \\ 
=\int_{0}^{\infty }\left\{ \ln \left[ \frac{f\left( t\right) }{f_{res}\left(
t\right) }\right] -\ln \left( r_{\exp}\mu \right) \right\} g\left( t\right) dt \\ 
\text{ } \\ 
=\int_{0}^{\infty }\ln \left[ \frac{f\left( t\right) }{f_{res}\left(
t\right) }\right] g\left( t\right) dt-\ln \left( r_{\exp}\mu \right)%
\end{array}%
\right.  \label{B134}
\end{equation}%
(using Eq. (\ref{B120}) and Eq. (\ref{A144}))%
\begin{equation}
\left. 
\begin{array}{l}
=\left[ D\left( g|f_{res}\right) -D\left( g|f\right) \right] -D\left(
f|f_{res}\right) \\ 
\text{ } \\ 
=D\left( g|f_{res}\right) -\left[ D\left( g|f\right) +D\left(
f|f_{res}\right) \right] \text{ .}%
\end{array}%
\right.  \label{B135}
\end{equation}%
So, combining Eq. (\ref{B131}) and Eqs. (\ref{B134})-(\ref{B135}) together
yields%
\begin{equation}
\int_{0}^{\infty }\left[ E\left( \tau \right) -\eta \right] F\left( \tau
\right) w\left( \tau \right) d\tau =D\left( g|f_{res}\right) -\left[ D\left(
g|f\right) +D\left( f|f_{res}\right) \right] \text{ .}  \label{B136}
\end{equation}

Now, setting $w\left( \tau \right) =\frac{1}{\mu }$ implies that $W\left(
t\right) =\frac{1}{\mu }t$, and hence%
\begin{equation}
g\left( t\right) =W\left( t\right) f\left( t\right) =\frac{1}{\mu }tf\left(
t\right) =f_{tot}\left( t\right) \text{ .}  \label{B137}
\end{equation}%
Substituting Eq. (\ref{B137}) into Eq. (\ref{B135}) yields Eq. (\ref{510}).

Setting $w\left( \tau \right) =n\left( n-1\right) F\left( \tau \right)
^{n-2}f\left( \tau \right) $ implies that $W\left( t\right) =nF\left(
t\right) ^{n-1}$, and hence%
\begin{equation}
g\left( t\right) =W\left( t\right) f\left( t\right) =nF\left( t\right)
^{n-1}f\left( t\right) =f_{\max }\left( t\right) \text{ .}  \label{B138}
\end{equation}%
Substituting Eq. (\ref{B138}) into Eq. (\ref{B135}) yields Eq. (\ref{520}).

\subsection{Derivation of Eq. (\protect\ref{530})}

Let $w\left( \tau \right) $ ($\tau >0$) be a positive-valued `weight
function', set $\bar{W}\left( t\right) =\int_{t}^{\infty }w\left( \tau
\right) d\tau $ ($t>0$), and further set $g\left( t\right) =\bar{W}\left(
t\right) f\left( t\right) $ ($t>0$). Multiplying both sides of Eq. (\ref{301}) by $%
F\left( \tau \right) w\left( \tau \right) $, and integrating over the
positive half-line, yields%
\begin{equation}
\left. 
\begin{array}{l}
\int_{0}^{\infty }\left[ E\left( \tau \right) -\eta \right] F\left( \tau
\right) w\left( \tau \right) d\tau \\ 
\text{ } \\ 
=\int_{0}^{\infty }\left\{ \int_{0}^{\tau }\ln \left[ \frac{r_{\exp}}{H\left(
t\right) }\right] f\left( t\right) dt\right\} w\left( \tau \right) d\tau \\ 
\text{ } \\ 
=\int_{0}^{\infty }\ln \left[ \frac{r_{\exp}}{H\left( t\right) }\right] f\left(
t\right) \left\{ \int_{t}^{\infty }w\left( \tau \right) d\tau \right\} dt \\ 
\text{ } \\ 
=\int_{0}^{\infty }\ln \left[ \frac{r_{\exp}}{H\left( t\right) }\right] f\left(
t\right) \bar{W}\left( t\right) dt \\ 
\text{ } \\ 
=-\int_{0}^{\infty }\ln \left[ \frac{H\left( t\right) }{r_{\exp}}\right] g\left(
t\right) dt%
\end{array}%
\right.  \label{B141}
\end{equation}%
(using Eq. (\ref{B134}))%
\begin{equation}
=\ln \left( r_{\exp} \mu \right) -\int_{0}^{\infty }\ln \left[ \frac{f\left(
t\right) }{f_{res}\left( t\right) }\right] g\left( t\right) dt  \label{B142}
\end{equation}%
(using Eq. (\ref{A144}))%
\begin{equation}
=D\left( f|f_{res}\right) -\int_{0}^{\infty }\ln \left[ \frac{f\left(
t\right) }{f_{res}\left( t\right) }\right] g\left( t\right) dt  \label{B143}
\end{equation}%
(considering $g\left( t\right) $ to be a density function and using Eq. (\ref%
{B120}))%
\begin{equation}
\left. 
\begin{array}{l}
=D\left( f|f_{res}\right) -\left[ D\left( g|f_{res}\right) -D\left(
g|f\right) \right] \\ 
\text{ } \\ 
=\left[ D\left( g|f\right) +D\left( f|f_{res}\right) \right] -D\left(
g|f_{res}\right) \text{ .}%
\end{array}%
\right.  \label{B144}
\end{equation}%
So, combining Eqs. (\ref{B141})-(\ref{B144}) together yields%
\begin{equation}
\int_{0}^{\infty }\left[ E\left( \tau \right) -\eta \right] F\left( \tau
\right) w\left( \tau \right) d\tau =\left[ D\left( g|f\right) +D\left(
f|f_{res}\right) \right] -D\left( g|f_{res}\right) \text{ .}  \label{B145}
\end{equation}

Now, setting $w\left( \tau \right) =n\left( n-1\right) \bar{F}\left( \tau
\right) ^{n-2}f\left( \tau \right) $ implies that $\bar{W}\left( t\right) =n%
\bar{F}\left( t\right) ^{n-1}$, and hence%
\begin{equation}
g\left( t\right) =\bar{W}\left( t\right) f\left( t\right) =n\bar{F}\left(
t\right) ^{n-1}f\left( t\right) =f_{\min }\left( t\right) \text{ .}
\label{B146}
\end{equation}%
Substituting Eq. (\ref{B146}) into Eq. (\ref{B145}) yields Eq. (\ref{530}).

\end{document}